# ARTICLE

# Thermomechanical properties of zero thermal expansion materials from theory and experiments

Andreas Erlebach,[a] Christian Thieme,[b] Carolin Müller,[a] Stefan Hoffmann,[a,d] Thomas Höche,[b] Christian Rüssel,[c] Marek Sierka*[a]

Origin and composition dependence of the anisotropic thermomechanical properties are elucidated for $Ba_{1-x}Sr_xZn_2Si_2O_7$ (BZS) solid solutions. The high-temperature phase of BZS shows negative thermal expansion (NTE) along one crystallographic axis and highly anisotropic elastic properties characterized by X-Ray diffraction experiments and simulations at the density functional theory (DFT) level. *Ab initio* molecular dynamics simulations provide accurate predictions of the anisotropic thermal expansion in excellent agreement with experimental observations. The NTE considerably decreases with increasing Sr content $x$. This is connected with the composition dependence of the vibrational density of states (VDOS) and the anisotropic Grüneisen parameters. The VDOS shifts to higher frequencies between 0-5 THz due to substitution of Ba with Sr. In the same frequency range, vibrational modes contributing most to the NTE are found. In addition, phonon calculations using the quasi-harmonic approximation revealed that the NTE is mainly connected with deformation of four-membered rings formed by $SiO_4$ and $ZnO_4$ tetrahedra. The thermomechanical and vibrational properties obtained in this work provide the basis for future studies facilitating the targeted design of BZS solid solutions as zero or negative thermal expansion material.

## Introduction

The vast majority of known substances expand with increasing temperature. However, there are a few crystals that contract in one or more crystallographic directions upon heating.[1–3] Among them are several silicates and aluminosilicates, in particular lithium alumosilicates with framework topologies of β-eucryptite, spodumene, β-quartz, and keatite.[4–6] Particularly important application of these compounds are zero thermal expansion (ZTE) materials, i.e., materials that barely or do not alter their volume in response to temperature. They find broad applications ranging from commonplace ceramic hobs through telescope mirrors to micromechanical devices. ZTE materials are usually synthesized as glass ceramics with a fairly high volume concentration of the phase with a negative coefficient of thermal expansion (CTE) and a residual glass phase which exhibits positive thermal expansion. In the past decade, numerous other materials such as $ZrMo_2O_8$, which is a metastable phase with a negative average CTE of around $-5 \cdot 10^{-6}$ $K^{-1}$, were found.[7] Other compounds with very low or even negative CTEs are $Zr_2P_2O_9$ (zirconyl phosphate) and $ZrV_2O_7$ (zirconium pyrovanadate) as well as $BaAl_2B_2O_7$.[8–10] The latter one can easily be crystallized from glasses, while the Zr-containing compounds cannot be precipitated from glasses in a high volume concentration. Unfortunately, $BaAl_2B_2O_7$ does not have the chemical durability required for many applications.

Recently, a new silicate has been reported that exhibits a negative coefficient of thermal expansion.[11] This compound, with the chemical formula $Ba_{0.6}Sr_{0.4}Zn_2Si_2O_7$, has an orthorhombic unit cell with the space group *Cmcm* and a crystal structure similar to the high temperature phase of $BaZn_2Si_2O_7$.[11] The CTE of $BaZn_2Si_2O_7$ is strongly negative in the direction of the crystallographic a-axis, while it is moderately positive in b- and c-direction.[12] It should be noted, that the axis with negative thermal expansion is, e.g., in Ref. 11 denoted as b-axis due to the change of crystal symmetry upon substitution of $Ba^{2+}$ with $Sr^{2+}$. The high-temperature (HT) phase of $BaZn_2Si_2O_7$ has also a negative CTE, however, below 280 °C, it transforms to a monoclinic low-temperature (LT) phase exhibiting a very high mean CTE of about $17.6 \cdot 10^{-6}$ $K^{-1}$.[12,13] The temperature range in which the low temperature phase is stable can be extended to higher temperatures by the partial (or total) replacement of Zn by other divalent transition metal ions such as Co, Mg, Ni, Cu, or Mn.[12,14,15] By contrast, the replacement of Ba by Sr shifts the temperature of the phase transition to lower values.[11,16] If Ba is substituted by Sr and simultaneously Zn by other divalent transition metals, in certain composition ranges also the high temperature phase is stable and negative CTE is observed.[17] Furthermore, Si can partially be replaced by Ge; the

[a.] *Otto Schott Institute of Materials Research, Friedrich Schiller University of Jena, Löbdergraben 32, 07743 Jena, Germany*

[b.] *Fraunhofer Institute for Microstructure of Materials and Systems IMWS, Walter-Hülse-Straße 1, 06120 Halle (Saale), Germany*

[c.] *Otto Schott Institute of Materials Research, Friedrich Schiller University of Jena, Fraunhoferstr. 6, 07743 Jena, Germany*

[d.] *Applied Systems Biology, Leibniz Institute for Natural Product Research and Infection Biology - Hans Knöll Institute, Adolf-Reichwein-Strasse 23, 07745 Jena, Germany*

\* corresponding author; marek.sierka@uni-jena.de

composition Ba$_{0.5}$Sr$_{0.5}$Zn$_2$SiGeO$_7$ has a negative CTE.[18] In analogy to lithium alumosilicates, all these solid solutions can be crystallized from glasses with appropriate compositions.

Although several studies on the formation of compounds based on BaZn$_2$Si$_2$O$_7$ and on the effect of the chemical composition on their thermal properties were reported recently, the number of possible elemental substitutions is very large and it is difficult to predict the CTE or even more the anisotropy of the thermal expansion.[11,14,18] The latter is an important parameter for the stresses in the glass ceramics formed during cooling from the synthesis temperature.

Therefore, detailed understanding and predictions of the anisotropic thermomechanical properties is of crucial importance for the targeted design of zero thermal expansion (ZTE) materials. In particular, atomistic simulations at the density functional theory (DFT) level proved to accurately predict the thermal expansion and elastic properties for a variety of materials.[19–21] A pivotal quantity for characterization of the thermal expansion mechanisms is the Grüneisen parameter that links the thermal and mechanical properties of a solid.[22] Furthermore, the microscopic Grüneisen parameter describes the volume dependency of the phonon frequencies. Its calculation using the quasi-harmonic approximation (QHA) allows the elucidation of vibrational modes that contribute to the negative thermal expansion (NTE).[21] However, accurate predictions of the structure and properties require explicit consideration of anharmonicity, e.g., using *ab initio* molecular dynamics (AIMD) simulations.[23] This applies particularly to materials showing pronounced anharmonic lattice vibrations and to predictions at elevated temperatures.[24]

This work combines DFT calculations and low-temperature X-ray diffraction experiments for elucidation of the anisotropic thermomechanical properties and the Grüneisen parameters for the HT phase of Ba$_{1-x}$Sr$_x$Zn$_2$Si$_2$O$_7$ (BZS) solid solutions with $x$ = 0.0, 0.25, 0.5, and 0.75. In addition, phonon calculations using the quasi-harmonic approximation reveal the origin of the negative thermal expansion. Finally, it is demonstrated that AIMD simulations facilitate accurate predictions of the anisotropic thermal expansion.

## Theory and computational details

**Modeling of anisotropic thermomechanical properties**

In general, the coefficient of thermal expansion $\alpha$ is related to the bulk modulus $K$, heat capacity $C_V$ and volume $V$ through the macroscopic Grüneisen parameter $\gamma$.[22,24] Here, the anisotropic thermal expansion $\boldsymbol{\alpha} = (\alpha_a, \alpha_b, \alpha_c)^\mathrm{T}$ for each crystallographic axis $a, b, c$ is considered due to the orthorhombic crystal structure of the HT BZS phase. Together with the three corresponding Grüneisen parameters $\boldsymbol{\gamma} = (\gamma_a, \gamma_b, \gamma_c)^\mathrm{T}$ and the (3×3) compliance tensor $\mathbf{S} = \mathbf{C}^{-1}$, which is inverse of the stiffness tensor $\mathbf{C}$, and neglecting shearing of the unit cell, $\boldsymbol{\alpha}$ is given as[25,26]

$$\boldsymbol{\alpha} = \frac{V}{C_V} \mathbf{S}\boldsymbol{\gamma}. \qquad (1)$$

In this work, $C_V$ and $\mathbf{S}$ are derived from DFT simulations. The Grüneisen parameters $\boldsymbol{\gamma}$ are obtained by fitting $\boldsymbol{\alpha}$ (eq 1) to experimentally determined cell parameters (this and previous works[13,16,17]) at temperatures $T$ from 100 to 800 K as well as DFT calculated cell parameters ($T$ = 0 K) of the HT phase for Ba$_{1-x}$Sr$_x$Zn$_2$Si$_2$O$_7$ solid solutions with $x$ = 0, 0.25, 0.5, and 0.75. The three adjustable parameters $\gamma_a, \gamma_b, \gamma_c$ and the compliance tensor $\mathbf{S}$ are assumed to be independent of temperature. In addition, using the DFT derived $\mathbf{C}$ and $\mathbf{S}$, elastic properties can be obtained such as the bulk modulus $K$ and the linear Young's modulus $Y_i$ along each crystallographic direction $i = a, b, c$.[27] Similarly, the linear compressibility $b_i$ for each crystallographic direction is calculated as[24,26]

$$b_i = \sum_j S_{ij}. \qquad (2)$$

The average (volumetric) Grüneisen parameter $\overline{\gamma}$ provides the relation between the volumetric thermal expansion $\alpha$, $K$, and $C_V$[26]

$$\overline{\gamma} = \frac{b_a\gamma_a + b_b\gamma_b + b_z\gamma_c}{b_a + b_b + b_c}. \qquad (3)$$

Elucidation of the microscopic origin of the (negative) thermal expansion at the atomic level used the microscopic Grüneisen parameters $\gamma_{\mathbf{k}i}$ for phonon branch $i$ and reciprocal lattice vector $\mathbf{k}$ obtained from phonon calculations (*cf.* Computational details section) along with the QHA. They quantify the volume dependence of the vibrational frequencies $\nu_{\mathbf{k}i}$ defined as[25,26]

$$\gamma_{\mathbf{k}i} = -\frac{V}{\nu_{\mathbf{k}i}}\frac{\partial \nu_{\mathbf{k}i}}{\partial V}. \qquad (4)$$

In addition, the DFT calculated $\gamma_{\mathbf{k}i}$ allow the determination of the temperature dependent, macroscopic Grüneisen parameter $\gamma_\mathrm{DFT}$[28]

$$\gamma_\mathrm{DFT} = \sum_{\mathbf{k}i} \frac{C_{\mathbf{k}i}\gamma_{\mathbf{k}i}}{C_V}, \qquad (5)$$

along with the contribution $C_{\mathbf{k}i}$ of each vibrational mode $\mathbf{k}i$ at $k$-point $\mathbf{k}$ for phonon branch $i$ to the heat capacity $C_V = \sum_{\mathbf{k}i} C_{\mathbf{k}i}$

$$C_{\mathbf{k}i} = ku^2\exp(u)(\exp(u) - 1)^{-2}, \qquad (6)$$

with $u = h\nu_{\mathbf{k}i}(kT)^{-1}$. Please note, the angular frequency $\omega$ was used in Ref. 28 for the quantity $u$ in contrast to the present work.

**Computational details**

Starting point for modelling of Ba$_{1-x}$Sr$_x$Zn$_2$Si$_2$O$_7$ solid solutions was the HT phase of BaZn$_2$Si$_2$O$_7$ ($x$ = 0) with the crystal symmetry *Cmcm* (space group No. 63).[12] Its unit cell contains four lattice positions for Ba$^{2+}$. In case of x = 0.25 and 0.75, there is only one symmetrically inequivalent crystal structure generated by replacing Ba$^{2+}$ with Sr$^{2+}$. For $x$ = 0.5, three symmetrically inequivalent initial structure models were constructed and remaining calculations used the lowest energy structure obtained. All initial structures were geometrically optimized using constant (zero) pressure conditions.

All DFT calculations used the *Vienna ab initio Simulation Package* (VASP)[29,30] along with the Projector Augmented Wave (PAW) method[31,32] and the PBEsol[33,34] exchange correlation functional. The empirical dispersion correction of Grimme *et al.* (D3)[35] was also added. Unless stated otherwise, calculations employed Monkhorst-Pack grids[36] with a *k*-point density of about 13 Å along every reciprocal lattice vector. The constant pressure optimizations and calculation of elastic constants used an energy cutoff of 900 eV for the plane wave basis sets, while phonon calculations used a cutoff of 400 eV.

Determination of the elastic constants[37] of the stiffness tensor **C** used the stress tensor calculated for six unit cell deformations and the stress-strain relation (Hookes law) as implemented in VASP.[29,30] Calculation of the vibrational density of states (VDOS) employed the finite difference (frozen-phonon) approach implemented in the program Phonopy[28] along with 2×1×2 supercells of the HT phase. The heat capacities $C_V$ (eq 1) were calculated from the phonon frequencies[38] using 11×11×11 Monkhorst-Pack grids[36] for the 2×1×2 supercells. Phonon dispersions show no imaginary frequencies (see supporting information) except for small imaginary frequencies (> -0.1 THz) of the acoustic modes close to the Γ-point. In order to shed a light on the atomistic origin of the zero thermal expansion, the microscopic Grüneisen parameters $\gamma_{\mathbf{k}i}$ (eq 4) were calculated for $x$ = 0.5. For this, phonon calculations were performed along with the QHA for isotropically deformed unit cells with a volume expansion and contraction of -3% and +3%, respectively. Calculation of the average volume expansion for $x$ = 0.5 used additional phonon calculations at volume strains of -9%, -6%, +1% and +3% along with fits of the energy-volume curves to the Vinet equation of state.[39,40]

Finally, the prediction of anisotropic thermal expansion used AIMD simulations for the HT phase of $BaZn_2Si_2O_7$ and $Ba_{0.5}Sr_{0.5}Zn_2Si_2O_7$. For this, the geometrically optimized unit cells were equilibrated for 15 ps at temperatures ranging from 200 to 1100 K (in 100 K steps) employing the isothermal-isobaric (NPT) ensemble along with the Langevin thermostat[41] and the Parrinello-Rahman barostat.[42] The friction coefficients for every atom and the lattice parameters were 2 ps$^{-1}$. During equilibration, the unit cell volume and lattice parameters were relaxed using a fictitious mass of 5 amu and a time step of 2 fs. Since the HT phase of $BaZn_2Si_2O_7$ transforms into the monoclinic LT phase[12] below 550 K, AIMD simulations with the same setup were applied to the LT phase ($x$ = 0) for temperatures between 200 and 600 K. In order to reduce the computational effort of AIMD simulations the softer PAW pseudopotential for oxygen was used along with an energy cutoff of 350 eV and 1×1×2 Monkhorst-Pack grids for the HT phase ($x$ = 0.0, 0.5), and only the Γ-point for the LT phase ($x$ = 0.0).

## Experimental Procedure

Samples with the compositions $Ba_{1-x}Sr_xZn_2Si_2O_7$ (with $x$ = 0.25, 0.5, and 0.75) were prepared from the following raw materials: $BaCO_3$, $SrCO_3$, ZnO, and $SiO_2$. From the raw materials, batches for 500 g of the final stoichiometric compositions were prepared. The compositions were melted in an induction furnace using a Pt crucible at a temperature of 1400 °C while stirring with a Pt-stirrer for 1.5 h. Then, the melt was poured into water. The sieved partially crystalline products were thermally treated at a temperature of 1250 °C kept for 1 h.

The temperature-dependent measurements were performed on a Stoe diffractometer using CuK$_\alpha$-radiation at temperatures of 103, 173 and 293 K in transmission using an $SiO_2$ glass capillary. The lattice constants were determined by Rietveld refinement.

## Results and discussion

### Structure and mechanical properties

Figure 1 shows XRD patterns of all samples studied in comparison to the theoretic pattern calculated from the ICSD data file 429938 of $Ba_{0.6}Sr_{0.4}Zn_2Si_2O_7$. All observed XRD lines are attributed to distinct lines from ICSD data. For $x$ = 0.5, the 2θ-values show the best agreement, while for $x$ = 0.25, the 2θ-values are shifted to lower values and for $x$ = 0.75 to higher values. This is due to the smaller ionic radius of $Sr^{2+}$ in comparison to $Ba^{2+}$ which results in smaller lattice constants and hence larger 2θ-values for increasing $Sr^{2+}$ concentrations. In Figure S1 (see supporting information), X-ray diffraction patterns of $Ba_{0.5}Sr_{0.5}Zn_2Si_2O_7$ at different temperatures of 103 K, 173 K, and 293 K are presented. The patterns have approximately the same shape and most notably, a phase transition to the low temperature phase does not take place, even at 103 K.

In Figure 2, a narrower 2θ-range from 26 to 40 ° is shown. The (022) peak at 30.4 ° is shifted to smaller 2θ-values with increasing temperatures which is attributed to larger lattice constants. By contrast, the (220)-peak observed at approximately 26.9° is shifted to larger 2θ values at higher temperatures which is related to decreasing lattice constants. This is in qualitative agreement with the behavior reported in Refs. 11,16 for temperatures above room temperature. Since a contraction of the lattice during heating occurs only in the direction of the crystallographic *a* axis, the lattice spacing increases for the (022)-peak, because it is not affected by the contraction of the *a*-axis, while the spacing (220) decreases because it strongly depends on the contraction of the *a* axis. The cell parameters obtained from Rietveld refinement and DFT calculations are summarized in Table S1 (see supporting information).

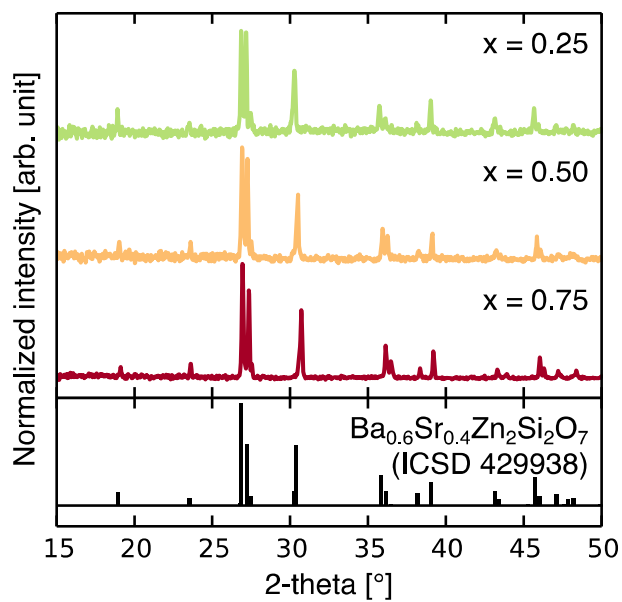

**Figure 1** XRD-patterns of the studied samples recorded at room temperature.

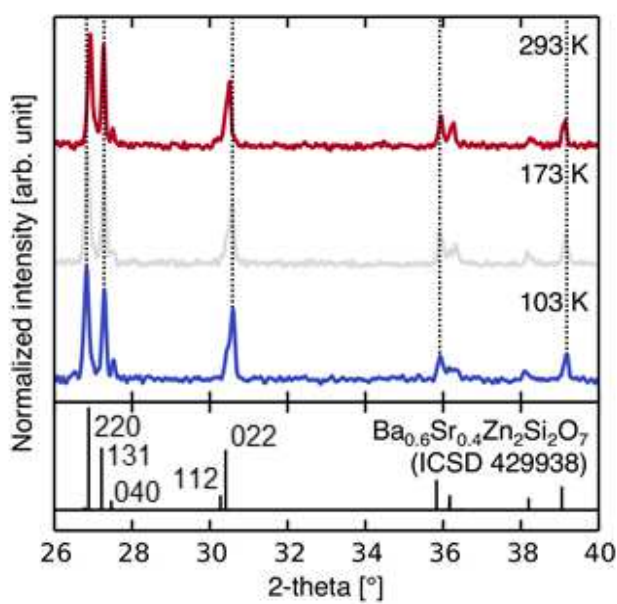

**Figure 2** X-ray diffractograms of $Ba_{0.5}Sr_{0.5}Zn_2Si_2O_7$ at different temperatures in a 2θ range from 26 to 40°. The dashed lines are just a guide for the eyes and run though the maximum of the most intense peaks measured at 103 K.

Table 1 shows the calculated elastic properties of $Ba_{1-x}Sr_xZn_2Si_2O_7$: the (Reuss) average bulk modulus $K$, Young's modulus $Y_i$ and the linear compressibilities $b_i$ along the crystallographic axes ($i = a, b, c$). In addition, the elements of the stiffness $\mathbf{C}$ as well as compliance $\mathbf{S} = \mathbf{C}^{-1}$ tensor are summarized in Table S2 (see supporting information). For all chemical compositions highly anisotropic elastic properties were predicted, with only weak dependence on the chemical composition. While the crystallographic $a$ axis shows the highest compressibility ($\mathbf{S}$ and $b_i$) or lowest stiffness ($\mathbf{C}$ and $Y_i$), respectively, the $b$ axis is the least compressible one. With the exception of $S_{bb}$ and $S_{cc}$, the compliance tensor $\mathbf{S}$ (and also $\mathbf{C}$) shows no systematic change for $x$ = 0.0, 0.25, and 0.5. Such small changes are expected to lie within the accuracy of the employed DFT method. This applies also the bulk modulus $K$. However, for $x$ = 0.75 a slightly higher value for $S_{aa}$ was obtained yielding higher compressibility $b_a$ of the crystallographic $a$ axis showing NTE. In contrast, the $b$ and $c$ axis show a systematic decrease of the compressibility with increasing $x$.

Together with the heat capacity at constant volume $C_V$ calculated using phonon calculations, the compliance tensor $\mathbf{S}$ shown in Table S2 and the experimentally observed cell parameters are used for determination of the anisotropic thermal expansion **α** (eq 1). Please note, calculations of vibrational properties used only the lowest energy structure (space group PMMA, no. 51) of the three possible Ba/Sr positions. The remaining Ba/Sr configurations are only 0.7 (space group PNNM, no. 59) and 1.3 meV/atom (space group AMM2, no. 38) higher in energy. Therefore, all configuration states show almost same probability that indicates random occupancy of Sr on Ba lattice sites in agreement with experimental observations.

**Table 1** Calculated linear compressibility $b_i$ [TPa$^{-1}$], Bulk (Reuss) and Young's moduli [GPa] of $Ba_{1-x}Sr_xZn_2Si_2O_7$.

| $x$ | $b_a$ | $b_b$ | $b_c$ | $K$ | $Y_a$ | $Y_b$ | $Y_c$ |
|---|---|---|---|---|---|---|---|
| 0.00 | 8.9 | 0.8 | 1.6 | 89 | 36 | 97 | 59 |
| 0.25 | 9.7 | 0.6 | 1.1 | 87 | 34 | 99 | 59 |
| 0.50 | 9.1 | 0.7 | 1.4 | 89 | 37 | 101 | 65 |
| 0.75 | 11.7 | 0.1 | 0.3 | 83 | 30 | 101 | 65 |

**Anisotropic thermal expansion**

Figure 3 shows the experimentally determined lattice parameters obtained in this and previous[13,16,17] studies along with the calculated DFT-values ($T$ = 0 K, crosses). In case of the HT phase of $BaZn_2Si_2O_7$ ($x$ = 0), experimentally determined cell parameters are only available above 550 K due to the martensitic phase transition.

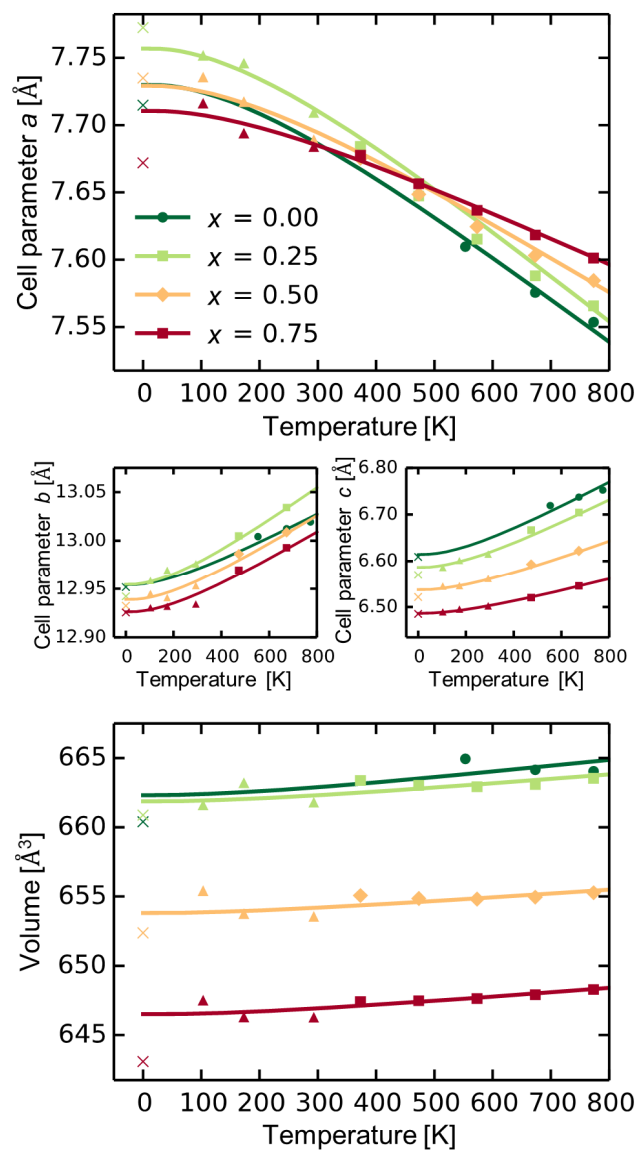

**Figure 3** Fit of the anisotropic thermal expansion **α** (solid lines, eq 1) to DFT calculated cell parameters ($T$ = 0 K, crosses) and experimentally determined lattice constants as a function of temperature for $Ba_{1-x}Sr_xZn_2Si_2O_7$ (triangles: this work; dots: ref. 13; squares: ref. 16; diamonds: ref. 17). Please note, all data points shown were included in the fit.

expansion that considerably deceases with increasing $Sr^{2+}$ content. At 300 K, its $\alpha_c$ is halved from 30×10$^{-6}$ K$^{-1}$ for $x$ = 0.0 to 15×10$^{-6}$ K$^{-1}$ for $x$ = 0.75. The value of $\alpha_b$ is virtually independent of the chemical composition with values ranging from 7×10$^{-6}$ to 10×10$^{-6}$ K$^{-1}$. The large anisotropy of both, elastic properties and thermal expansion, can lead to cracks and residual stresses in the microstructure of the glass ceramics depending on orientation of the crystallites. For example, in case of surface crystallization highly directional crystallite orientations were observed leading to undesired crack formation perpendicular to the $c$ axis.[43] Therefore, knowledge of the anisotropic elastic properties (Tab. 1) and the CTE as well as their dependence on the chemical composition provide vital input for continuum mechanics simulations, *e.g.*, using the finite element method,[44,45] for the targeted design of suitable microstructures to facilitate ZTE in the desired temperature range.

The fit of **α** to the cell parameters (Fig. 3) provides the linear Grüneisen parameters $\gamma_i$ that are the adjustable parameters of the model function (*cf.* eq 1). Table 2 shows the resulting linear and average Grüneisen parameters $\bar{\gamma}$, which are assumed to be independent of temperature. The Grüneisen parameters for $x$ = 0.0 show no clear correlation with the remaining ones, since the cell parameters of HT phase are only available for temperatures above 550 K resulting in erroneous Grüneisen parameters that are assumed to be temperature independent. For $x$ = 0.25, 0.5 and 0.75, $\bar{\gamma}$ and $\gamma_a$ are almost zero. On the other hand, $\gamma_c$ is 0.23 for $x$ = 0.25 and decreases with increasing $Sr^{2+}$ concentration to 0.15 ($x$ = 0.75), while $\gamma_b$ is slightly lower or equal to $\gamma_c$. Due to the small differences between $\gamma_b$ and $\gamma_c$ the considerably lower thermal expansion of the $b$ axis compared to the $c$ axis is connected with its lower linear compressibility $b_b$. As can be seen from eq. 1, the NTE along the $a$ axis is $\alpha_a \propto (\gamma_a S_{aa} + \gamma_b S_{ab} + \gamma_c S_{ac})$. Since the compliances $S_{ij}$ (Tab. S2) show no significant change with the chemical composition, the decrease of $\gamma_b$ and $\gamma_c$ with increasing $x$ leads to the reduction of NTE of the crystallographic $a$ axis. This indicates that the change of the NTE along the $a$ axis is connected with the variation of vibrational states of the HT phase by substitution of $Ba^{2+}$ with $Sr^{2+}$.

**Table 2** Linear compressibility $b_i$ [TPa$^{-1}$] at the DFT level as well as linear $\gamma_i$ and average $\bar{\gamma}$ Grüneisen parameter obtained from least square fit of **α** (eq 1) to experimentally determined cell parameters of $Ba_{1-x}Sr_xZn_2Si_2O_7$ (Fig. 3).

| $x$ | $\gamma_a$ | $\gamma_b$ | $\gamma_c$ | $\bar{\gamma}$ |
|---|---|---|---|---|
| 0.00 | -0.16 | 0.02 | 0.10 | -0.11 |
| 0.25 | 0.01 | 0.18 | 0.23 | 0.04 |
| 0.50 | 0.01 | 0.15 | 0.18 | 0.03 |
| 0.75 | 0.03 | 0.15 | 0.15 | 0.04 |

The NTE of the crystallographic $a$ axis decreases with increasing $Sr^{2+}$ concentration, with $\alpha_a$ at 300 K increasing from -34×10$^{-6}$ K$^{-1}$ for $x$ = 0.25 (-32×10$^{-6}$ K$^{-1}$ for $x$ = 0.0) up to -19×10$^{-6}$ K$^{-1}$ for $x$ = 0.75. In contrast, the $c$ axis shows pronounced positive thermal

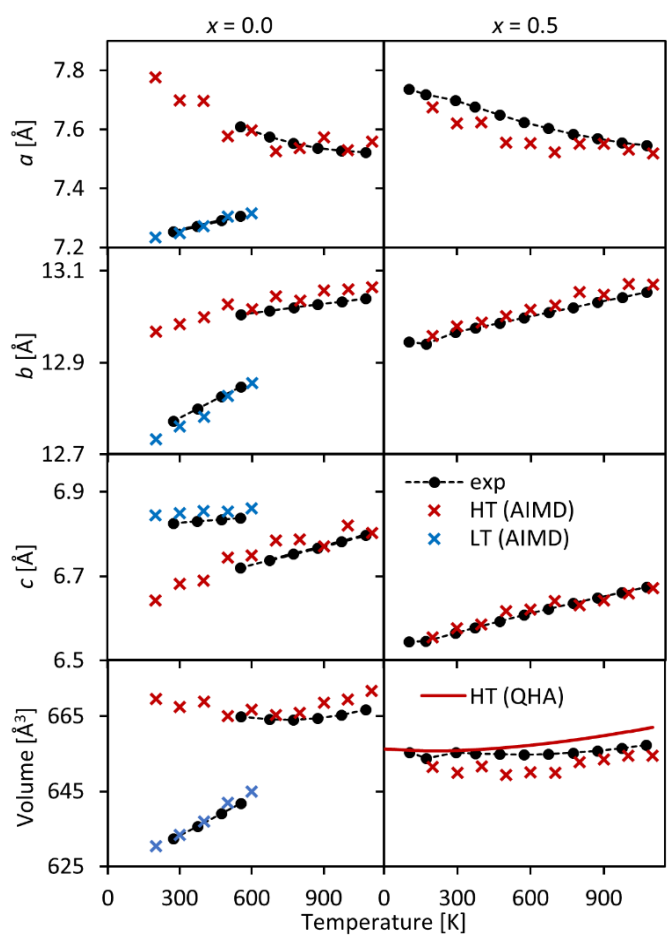

**Figure 4** Cell parameters and unit cell volume as a function of temperature of BaZn$_2$Si$_2$O$_7$ ($x$ = 0.0) and Ba$_{0.5}$Sr$_{0.5}$Zn$_2$Si$_2$O$_7$ ($x$ = 0.5) obtained from experiments (exp), *ab initio* molecular dynamics (AIMD) simulations and phonon calculations using the quasi-harmonic approximation (QHA).

Figure 4 compares the lattice parameters and unit cell volumes observed in experiments (this and previous studies) and calculated using AIMD simulations and QHA phonon calculations. Excellent agreement of the AIMD and experimental results were obtained, for both, the LT and HT phase of BaZn$_2$Si$_2$O$_7$ ($x$ = 0.0) as well as for the HT phase of Ba$_{0.5}$Sr$_{0.5}$Zn$_2$Si$_2$O$_7$ ($x$ = 0.5). This clearly proves the capability of the employed computational methodology not only to accurately predict the anisotropic thermal expansion but also to describe the anharmonic lattice vibrations of Ba$_{1-x}$Sr$_x$Zn$_2$Si$_2$O$_7$ solid solutions. In addition, phonon calculations using the QHA show similar accuracy for predictions of the volumetric thermal expansion up to 700 K compared to AIMD that explicitly consider the anharmonicity.

**Microscopic mechanism of the negative thermal expansion**

Figure 5 shows the harmonic vibrational density of states (VDOS) along with the partial VDOS for all elements of BaZn$_2$Si$_2$O$_7$ ($x$ = 0.0). In addition, the VDOS for Sr$^{2+}$ and Ba$^{2+}$ are shown for $x$ = 0.25, 0.5, and 0.75. The partial VDOS of Sr$^{2+}$ and Ba$^{2+}$, respectively, range from 0 to 7 THz in each case. While the Ba$^{2+}$ VDOS shows its maximum at about 1.5 THz, the maximum value of the Sr$^{2+}$ VDOS is approximately at 3 THz. Due to the lower mass of Sr$^{2+}$ the VDOS is clearly shifted towards higher frequencies with increasing Sr$^{2+}$ concentration $x$. In contrast, the partial VDOS of Zn$^{2+}$, Si$^{4+}$, and O$^{2-}$ are almost independent of chemical compositions showing only minor shifts to higher frequencies with increasing $x$ (not shown).

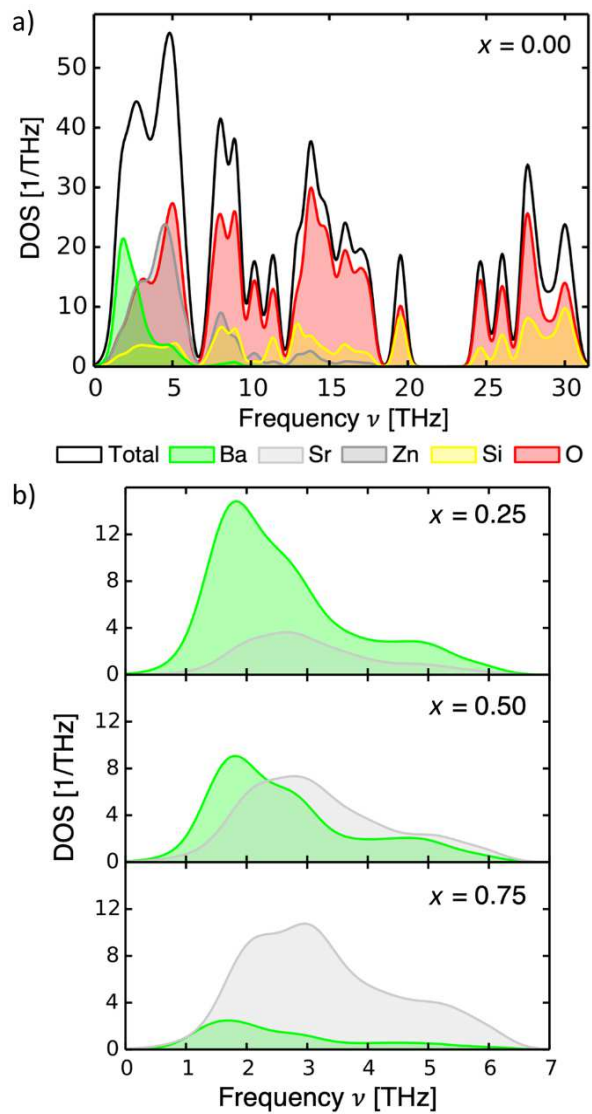

**Figure 5** Total and partial vibrational density of states (VDOS) of Ba$_{1-x}$Sr$_x$Zn$_2$Si$_2$O$_7$ solid solution for a) $x$ = 0 and b) partial VDOS of Ba$^{2+}$ and Sr$^{2+}$ as a function of the chemical composition $x$.

In order to shed a light on the microscopic mechanisms of the NTE, additional phonon calculations using the QHA were performed for $x$ = 0.5. This provided the microscopic (mode) Grüneisen parameters $\gamma_{\mathbf{k}i}$ for each branch $i$ at reciprocal lattice vector $\mathbf{k}$. Figure 6 depicts the distribution of the average (macroscopic) $\gamma_{\text{DFT}}$ (eq 5) calculated for frequency intervals of 1 THz at 300 K as well as number density $g(\gamma_{\mathbf{k}i})$ of the mode Grüneisen parameter $\gamma_{\mathbf{k}i}$. Negative (average) $\gamma_{\text{DFT}}$ were obtained at low frequencies between 0 to 5 THz along with a minimum value of -2.4 at 1.5 THz, that is the corresponding vibrational modes contribute most to the NTE. In the same frequency range, the main contributions of the $Ba^{2+}$ VDOS (maximum at about 1.5 THz) to the total VDOS are located, which is shifted towards higher frequencies when $Ba^{2+}$ is replaced by $Sr^{2+}$ (Fig. 5). For higher frequencies, $\gamma_{DFT}$ is positive or close to zero. This rationalizes the reduction of the NTE along $a$ direction with increasing $Sr^{2+}$ concentration (Fig. 3). The majority of the vibrational modes show positive Grüneisen parameter $\gamma_{\mathbf{k}i}$ as can be seen from $g(\gamma_{\mathbf{k}i})$ with its maximum value is at 0.5. Therefore, $\gamma_{DFT}$ is negative at low temperatures turning positive due to thermal excitation of higher frequency modes with increasing temperature showing mostly positive $\gamma_{\mathbf{k}i}$.

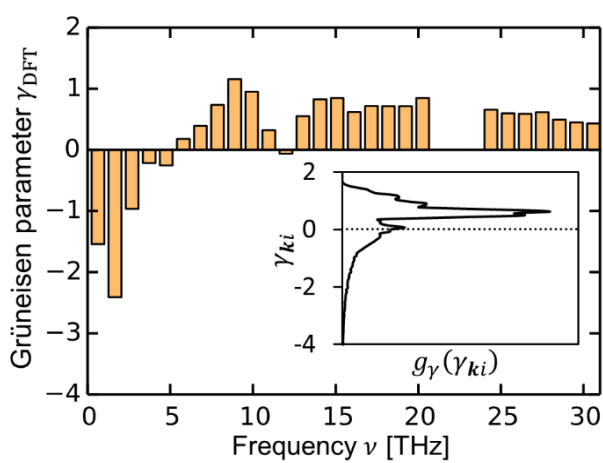

**Figure 6** Distribution of the average mode Grüneisen parameter $\gamma_{DFT}$ (eq 5) of $Ba_{0.5}Sr_{0.5}Zn_2Si_2O_7$ ($x$ = 0.5) as a function of vibrational frequency $\nu$ calculated using frequency intervals of 1 THz at $T$ = 300 K. The inset shows the number density of the mode Grüneisen parameter $\gamma_{\mathbf{k}i}$.

Atomic displacements along the polarization vector of the vibrational mode showing the most negative Grüneisen parameter $\gamma_{\mathbf{k}i}$ of -12.2 at a frequency of 1.3 THz are displayed in Figure 7a. A simplified sketch of the displacements of $Si^{4+}$ and $Zn^{2+}$ is depicted in Fig. 7b. In addition, the supporting information contains an animation of the vibrational mode. Basic building blocks of the crystal structure of the HT phase are $SiO_4$- and $ZnO_4$-tetrahedra. They form four-membered OSi–O–ZnO (**SOZ**) rings, which in turn form two sheets. Two bridging oxygen ions connect both sheets by Si–O–Si bridges, while $Ba^{2+}$ and $Sr^{2+}$ ions are located between the sheets. The vibrational mode shows an oscillation in direction of the $b$ axis in case of $Ba^{2+}$ and $Sr^{2+}$, respectively, as well as a rocking movement of the rigid Si–O–Si bridges within the $a$-$b$ plane. The latter leads to deformation of the **SOZ** rings yielding contraction of the $a$ axis. Similarly, previous experimental studies using high temperature X-ray diffraction also concluded that the deformations of $ZnO_4$- and $SiO_4$-tetrahedra lead to the NTE of the $a$ axis.[11]

In order to qualitatively evaluate the effect of this vibrational mode on the crystal lattice, the stress tensor of the turning point **+1** with (arbitrarily chosen) atomic displacements of about 2 Å along its polarization vector at the equilibrium (zero pressure) cell volume was calculated. This yielded negative normal stress in $a$ direction of about -2.8 GPa indicating that this mode clearly contributes to NTE along the $a$ axis. By contrast, in $b$ and $c$ direction, a normal stress of -0.02 and 2.4 GPa was obtained, respectively, showing that the vibrational mode has almost no effect on the $b$ axis and contributes to the relatively large thermal expansion $\alpha_c$.

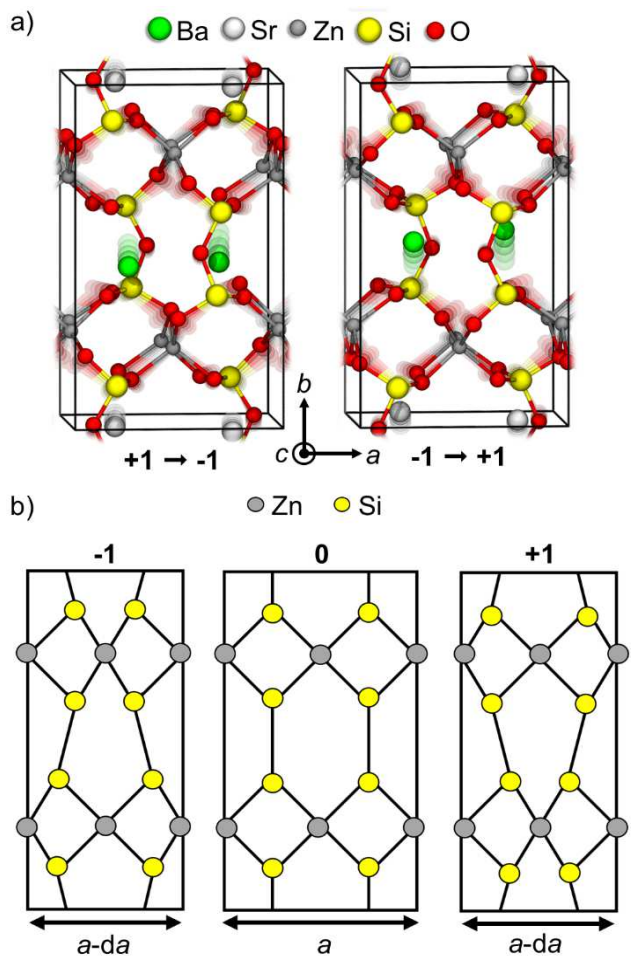

**Figure 7** Representation of the vibrational mode showing the most negative mode Grüneisen parameter $\gamma_{\mathbf{k}i}$. a) Atomic displacements along its polarization vector between the turning points **-1** and **+1** and (b) simplified representation of the displacements of Zn and Si compared to their equilibrium position (**0**).

## Conclusions

In summary, the mechanism and the composition dependence of the anisotropic thermomechanical properties have been investigated for $Ba_{1-x}Sr_xZn_2Si_2O_7$ at the atomic level. Predictions of the highly anisotropic thermal expansion using *ab initio* molecular dynamics simulations are in excellent agreement with experimental observations. In addition, the thermal expansion strongly depends on the chemical composition $x$. By contrast, the highly anisotropic elastic constants are almost independent of the chemical composition. Therefore, the change of the negative thermal expansion along the crystallographic $a$ axis has been attributed to the dependence of the Grüneisen parameters on the chemical composition, which quantify the volume dependence of lattice vibrations.

Largest changes of the VDOS have been obtained in the low frequency range (0-5 THz) at which the microscopic Grüneisen parameters show negative values. The considerable shift of VDOS in this frequency range due to the substitution $Ba^{2+}$ with $Sr^{2+}$ rationalizes that NTE along the *a* axis gets less negative with increasing *x*. Furthermore, this NTE originates from the deformation of four-membered OSi–O–ZnO rings as indicated by the calculated microscopic Grüneisen parameters. Characterization of the highly anisotropic thermal expansion and calculation of the elastic constants provides vital input for future simulations of size, shape and orientation of crystallites in the glass ceramics microstructure by employing continuum mechanics simulations. Such simulations, e.g., using the finite element method are capable to facilitate the targeted design of the microstructure of BZS glass ceramics with zero thermal expansion. Furthermore, additional phonon and molecular dynamics simulations can provide further insights of the thermomechanical properties and the martensitic phase transition of BZS with partial substitution of Si and Zn with other elements such as Ge or Mg, respectively.

## Conflicts of interest

There are no conflicts to declare.

## Acknowledgements

This work was funded by the German Federal Ministry of Education and Research under the Grant nos. 03VP01701 and 03VP01702.